\documentclass[11pt]{article}
\def\ie{i.e.}
\def\eg{ e.g.}
\def\e{\rm e}
\def\etal{et al.}
\begin{document}
\title{Shadowing of Virtual Photons in Nuclei\\ at Small $x_{\rm Bj}$ in the QCD
Dipole Picture\thanks{To be published in Acta Physica Polonica B, no. 3 (1998).}}
\author{A. Bialas and W. Czyz\\
Institute of Physics, Jagellonian University\\
Reymonta 4, 30-059 Krak\'ow, Poland}
\date{October 7, 1997}
\maketitle
\begin{abstract}
Compact and well defined formulae for the shadow of the virtual photon
interacting with a large nucleus at small $x_{\rm Bj}$ are given in the QCD dipole
picture. Two classes of contributions are considered: (a) quasi-elastic
interaction of the $q\bar{q}$ dipole and (b) multi-pomeron coupling.
\end{abstract}
\section {Introduction}

We have now at our disposal the QCD dipole picture of virtual photon
interactions \cite{mu1, mu2, mu3} with hadrons \cite{na1, na2} which
is able to account for existing data on proton structure function $F_2$ (\ie\ 
on total virtual photon--proton cross-section) at small $x_{\rm Bj}$. It is
therefore interesting to investigate whether the model can be also applied to
nuclear targets, in particular  if it can describe the shadowing of the virtual
photons. This becomes especially interesting in view of a possible extention of
the HERA experimental program to nuclear targets \cite{ar1}.

In our recent paper \cite{bi1} we have calculated in this model the double
scattering contribution to the virtual photon shadowing\footnote{The subject of photon shadowing in nuclei has a vast
literature. For recent reviews, see \eg\ \cite{ar1} and \cite{st1}.}.
The multiple scattering corrections, which are more difficult to handle, were
merely estimated, however. As the resulting shadowing at small $x_{\rm Bj}$ turned out to be rather
substantial and not too far 
from experimental data, it seems worthwhile to pursue the problem
further and to evaluate more precisely the multiple scattering terms. This is
the purpose of the present paper.
As in \cite{bi1} ({\it c.f.} also \cite{bi2, bi3}) we are dealing
with two types of contributions:

(a) The ``quasi-elastic'' interaction of the $q\bar{q}$ dipole (first introduced
in the dipole picture in \cite{bi3}) contributing mostly at small mass excitations
of the virtual photon.

(b) The ``direct'' interaction of the dipole cascade (corresponding to the
so-called multiple pomeron coupling in the Regge language) which dominates the
high-mass excitations.

The amplitudes describing these processes were derived and discussed in
\cite{mu1, mu2, mu3, bi2, bi3, bi4}.
When applied to the interaction with nuclear targets, however, they must be
modified in two important points:

{\it (i)} a phase difference related to the change in the longitudinal momentum of
system travelling through the nucleus must be taken into account. Consequently,
 at some point it is nessesary to transform the amplitudes   into the
momentum
representation. This makes the calculation substantially more complicated.

{\it (ii)} Since the transverse size of the gluon cascade is much smaller than that
of a large nucleus, we can use the standard approximation and integrate
 over the impact-parameter dependence of the elementary amplitudes, \ie\ 
 we need  to deal with forward scattering amplitudes only. Indeed, the
convolution of the onium and the nucleus transverse profiles is
approximately
\begin{equation}
\int T_A\, (\vec s)\, T (\vec b -\vec s)\,d^2\!s\approx
T_A\,(b)\int T\,(b')\,d^2b'=\tilde T(k=0)\,T_A\,(b)\,,
\label{aia}
\end{equation}
where $\tilde{T}(k=0)$ is the forward scattering amplitude and $T_A (b)$ is the nucleus profile.

 This simplifies the
calculation in an essential way.
Our ultimate goal is to give an explicit expression for the shadow, defined as
\begin{equation}
C(A,x_{\rm Bj},Q^2) = \frac{\sigma_{\rm tot}(A)}{A\sigma_{\rm tot}(n)} -1,   \label{i}
\end{equation}
where $\sigma_{\rm tot}(A)$ is the total photon-nucleus cross-section and
$\sigma_{\rm tot}(n)$ is the total photon-nucleon cross-section. $\sigma_{\rm tot}(n)$
has been measured \cite{he1} and successfully computed in the
dipole picture \cite{na1}. This will be our input.

To compute  $\sigma_{\rm tot}(A)$  we have to know the forward amplitudes for
(a) quasi-elastic scattering of the dipole and (b) direct scattering of the
dipole cascade from a given number, $N$, of nucleons. We then dress them up with
the light-cone photon wave functions \cite{bj1} and with single nucleon
densities characterizing the nuclear shape. Finally we add
them up for $2 \leq N \leq A$. This calculation is described in the next two
sections.
In Sec.~4 we discuss the obtained results and formulate our conclusions.

\section{Forward $\gamma^*$-nucleus amplitude from the direct interaction\\ of
the dipole cascade}
We start by giving the final result for the forward amplitude for scattering on $N$ nucleons of
the dipole cascade emerging from an onium. It reads
\begin{eqnarray}
&&F_N^{\rm dir}(r,r_0;Y,y) = (2N-3)!! \pi \alpha^2(\alpha^3N_c)^{N-1} r_0^{2N}
n_{\rm eff}^N
\e^{i(z_1-z_N) m x_P}
\nonumber \\ 
&&\times\prod_{k=1}^{N}\left(\int\limits_{c_k-i\infty}^{c_k+i\infty}
\frac{d\lambda_k}{2\pi i} h(\lambda_k) \e^{{\mit\Delta}(\lambda_k)y} G(\lambda_k)
\right) \e^{{\mit\Delta}(\gamma)(Y-y)} \left(\frac{r}{r_0}\right)^{\gamma} \frac{1}
{G(\gamma)},               \label{ii}
\end{eqnarray}
where
\begin{equation}
{\mit\Delta}(\lambda)= \frac{\alpha N_c}{\pi} \chi(\lambda) ,  \; \chi(\lambda)=
2\psi(1) -\psi\left(1-\textstyle{\frac{\lambda}{2}}\right) -\psi\textstyle({\frac{\lambda}{2}})\,,       \label{iii}
\end{equation}
$\alpha$ is the strong coupling constant, $N_{\rm c}$ is the number of colours,
$r_0$ is a parameter of the order of the effective size of the nucleon, $r$ is
the transverse size of the incident onium. The remaining symbols denote:
\begin{eqnarray}
h(\lambda)=\frac{4}{\lambda^2(2-\lambda)^2}\,, \;\;
G(\lambda)=\frac{{\mit\Gamma}\textstyle({\frac{\lambda}{2})}}{{\mit\Gamma}(1-\textstyle{\frac{\lambda}{2}})}\,, \;\;
\gamma=\sum_{k=1}^N\lambda_k\,,
\nonumber \\Y=-\log\textstyle{\frac{x_{\rm Bj}}{c}}\,,\;\;  y=-\log\textstyle{\frac{x_P}{c}}\,, \;\; x_P=
\frac{x_{\rm Bj}}{\beta}\,,
\;\; \beta=\frac{Q^2}{Q^2+M^2},       \label{iv}
\end{eqnarray}
and $M$ is the mass of the dipole cascade; $c$ is a constant giving
the energy scale, $n_{\rm eff}$ is an effective number of onia in one
nucleon (estimated in \cite{na1},\break {\it c.f.} also
\cite{bi1}).

The origin of the phase factor $\e^{i(z_1-z_N) m x_P}
=\e^{i(z_1-z_N)\frac{Q^2+M^2}{2\nu}}$
 where
$z_1$ and
$z_N$ are the
longitunal positions of the first and the last scattering\break ($m$ is nucleon mass
and $\nu$ is the energy of the photon ) was discussed in detail already by
Gottfried and Yennie \cite{go1}\footnote{When only one mass is excited, as it
is in the present case, the phase factor depends only on the first and last
points of the interaction.}.

The starting point in the derivation of Eq.~(\ref{ii}) is the generalization of
the Mueller--Patel formula \cite{mu2} for triple-pomeron coupling. In the case
of $N$ nucleons it reads
\begin{eqnarray}
&&F_N^{\rm dir}(r,r_1,\ldots r_N;Y,y)\nonumber \\
&&= \e^{i(z_1-z_N) m x_P}
\int\limits\left(\prod_{k=1}^N
\frac{dx_k}{x_k}\frac{dx'_k}{x'_k}
\tau(x_k,x'_k)n_1(r_k,x_k,y^*)\right)\nonumber \\
&&\times n_N(r,x'_1,\ldots ,x'_N;Y\!-\!y^*,y\!-\!y^*).
\label{v}
\end{eqnarray}
Here
\begin{equation}
n_1(r_k, x_k,y^*) =\int\limits_{c-i\infty}^{c+i\infty}\frac{d\lambda}{2\pi i}
\e^{{\mit\Delta}(\lambda)y^*}
\left(\frac{r_k}{x_k}\right)^{\lambda}      \label{vi}
\end{equation}
are  the densities of the dipoles of   the transverse size $x_k$ {\it integrated over
the trasverse positions of the dipoles}  within an onium of size
$r_k$ moving with rapidity $y^*$. $n_N(r,x_1,\ldots ,x_n; Y-y^*,y-y^*)$ is the
density of $N$ dipoles of the transverse sizes $x_1,\ldots x_N$ again {\it
integrated over the transverse positions of the dipoles} within the cascade of
the transverse size $r$ evolved from the incident onium. $\tau(x,x')$ is the
forward scattering amplitude of two dipoles of transverse sizes $x$ and $x'$
(in the two-gluon exchange approximation it is energy independent \cite{mu2}).
The arguments $r_1,\ldots ,r_N$ are the transverse sizes of the onia representing
the target nucleons. $Y,y$ are defined above in Eq.~(\ref{iv}). In the
terminology employing the pomeron concept, $y$ is the rapidity which divides
the process into the region of one pomeron (rapidities between $Y$ and $y$)
 and $N$ pomerons (rapidities smaller than $y$) in the $N+1$ pomeron coupling.

   To obtain a formula for $n_N(r,x_1,\ldots ,x_n; Y-y^*,y-y^*)$ one has to solve
an integro-diferential equation which one obtains by $N$-fold differentiation of
the generating functional given in \cite{mu3}. We do this by the method
described in \cite{bi3} for double-dipole scattering. An extension of this
procedure to arbitrary $N$ is described in Appendix A. At this point it should
be
emphasized that a solution in compact form, necessary to derive the result of
Eq.~(\ref{ii}), can be obtained only for distributions integrated over the
transverse position of the dipoles with respect to the original onium. The
solution for arbitrary dipole position is much more involved. Fortunately,
as already mentioned in the Introduction, for large nuclei the transverse size
of the
incident onium is much smaller than the nuclear diameter and thus  we only need
the distributions integrated over the transverse positions inside the onium.

In order to obtain the contribution to the shadow (Eq.~(\ref{i})) one has to add
the contributions from $N=2,3,\ldots A$ nucleons and average them over the positions
of the nucleon inside the nucleus and over the virtual photon wave functions.
Finally we have to integrate over the excited mass $M$\break (\ie\  the rapidity $y$).
Using (\ref{i}) and
\begin{equation}
\sigma_{\rm tot} = 2 {\rm Re} F    \label{vii}
\end{equation}
one obtains
\begin{eqnarray}
&&C^{\rm dir}(A,x_{\rm Bj},Q^2) =  \frac{-2}{A\sigma_{\rm tot}(n)}    \nonumber \\
&&\times {\rm Re}\left(\int\limits_{0}^{Y}
dy <\sum_{N=2}^A(-1)^N \frac{A!}{(A-N)!} n_{\rm eff}^N
 \langle\langle\psi_Q\mid F_N^{\rm dir}\mid \psi_Q\rangle \rangle \right) .
\label{viii}
\end{eqnarray}
The average    $ \langle\psi_Q\mid F_N^{\rm dir}\mid \psi_Q\rangle$ is the
integral of $F_N^{\rm dir}$ multiplied by light-cone photon densities \cite{bj1} given
by
\begin{eqnarray}
\mid {\mit\Psi}_Q(r,\eta)\mid^2 &=&\frac{N_c \alpha_{\rm em} e_{\rm f}^2}{\pi^2}
W(r,\eta,Q),
\label{ix}\\
W^{\rm T}(r,\eta,Q)&=& \textstyle{\frac12} [\eta^2+(1-\eta)^2] \hat{Q}^2 K_1^2(\hat{Q}r),
\label{ixa}\\
W^{\rm L}(r,\eta,Q)&=& 2 \eta(1-\eta)\hat{Q}^2 K_0^2(\hat{Q}r),
\label{ixb}
\end{eqnarray}
where $\hat{Q} = [\eta(1-\eta)]^{\frac12}Q, \alpha_{\rm em} =\frac1{137},
e_{\rm f}^2$ is the sum of the squares of the quark charges.
$\eta$ is the light-cone momentum fraction of one of the quarks in the photon.

The relevant integrals $\int\limits d^2rd\eta\ldots$ can be found in \cite{gr1} and they
result in the following prescription: $r^{\gamma} $ in (\ref{ii}) should be
replaced by the  expressions
\begin{equation}
r^{\gamma} \rightarrow\frac{N_c \alpha_{\rm em} e_{\rm f}^2}{\pi^2}
\frac{{\mit\Gamma}^2(2-\frac{\gamma}{2}){\mit\Gamma}^4(1+\frac{\gamma}{2})}{{\mit\Gamma}(4-\gamma){\mit\Gamma}(2+\gamma)}
\left(\frac{Q}{2}\right)^{-\gamma} I^{\rm T,L}(\gamma),    \label{x}
\end{equation}
where
\begin{equation}
I^{\rm T}(\gamma) =\frac{( 2-{\gamma\over 2})(1+\frac{\gamma}{2})}{{(\frac{\gamma}{2})}(1-\frac{\gamma}{2})},  \label{xa}
\end{equation}
\begin{equation}
 I^{\rm L}(\gamma) = 2 \;\;. \label{xb}
\end{equation}
Averaging over the nucleon positions reduces to multiplying
 $\langle\psi_Q\mid F_N^{\rm dir}\mid \psi_Q\rangle $ by $N$ single particle nucleon densities
$\rho(b,z_k)$ and integrating over $z_k$'s kept ordered:
\begin{eqnarray}
&&\langle\langle\psi_Q\mid F_N^{\rm dir}\mid \psi_Q\rangle\rangle = \langle\psi_Q\mid F_N^{\rm dir}\mid \psi_Q\rangle 
\nonumber
\\
&&\times\int\limits d^2b
\int\limits_{-\infty}^{+\infty}dz_1\int\limits_{z_1}^{+\infty} dz_N \e^{i(z_1-z_N)mx_P}
\rho(b,z_1)\rho(b,z_N) \nonumber \\
&&\times\int\limits_{z_1}^{z_N}dz_2 \rho(b,z_2)\ldots  \int\limits_{z_{N-2}}^{z_N}dz_{N-1}
\rho(b,z_{N-1})   \nonumber \\
&&= \langle\psi_Q\mid F_N^{\rm dir}\mid \psi_Q\rangle  \int\limits d^2b
\int\limits_{-\infty}^{+\infty}dz_1\int\limits_{z_1}^{+\infty} dz_N \e^{i(z_1-z_N)mx_P}
\rho(b,z_1)\rho(b,z_N) \nonumber \\ 
&&\times\frac1{(N-2)!} \left(\int\limits_{z_1}^{z_N}
dz \rho(b,z) \right)^{N-2}.          \label{xi}
\end{eqnarray}
Inserting (\ref{xi}) into  (\ref{viii}) we obtain the required formula for the
shadow.

For the sake of completeness let us also give the explicit formula for
$\sigma_{\rm tot} (n)$:
\begin{equation}
\sigma_{\rm tot} (n)=\sigma (r_0, Y)=2\pi\alpha^2 r_0^2 n_{\rm eff}
\int\frac{d\gamma}{2\pi i}
\left(\frac{r}{r_0}\right)^\gamma\e^{{\mit\Delta}(\gamma)y}h(\gamma)\,,
\end{equation}
where $r^\gamma$ is given by Eq.~(\ref{x}).

\section{Forward $\gamma^*$-nucleus amplitude from the quasi-elastic scattering
of a QCD dipole in the nucleus}

As is well-known since the seminal paper of Stodolsky \cite{s+1}
and of Gottfried and Yennie \cite{go1},
the multiple elastic scattering of a photon fluctuation inside a nucleus
contributes in an essential way to the nuclear shadowing. In the QCD dipole
picture this corresponds to quasi-elastic scattering of an onium whose
transverse
size is distributed according to the light-cone wave functions of the photon
given in \cite{bj1}. In the present section we derive the formula for this
contribution to the shadow.

The $N$-fold scattering contribution to the forward $\gamma^*$-nucleus
amplitude is built as a product of $N$ onium-nucleon  amplitudes (including the
eikonal phases) sandwiched
between the initial and final state of the virtual photon and summed over all
intermediate states of the onium as follows
\begin{eqnarray}
&&\langle\psi_Q\mid F_N^{\rm qel}\mid \psi_Q\rangle =n_{\rm eff}^N\langle Q\mid T(z_N) \mid k_{N-1}\rangle 
\nonumber\\
&&\times \frac{d^2 k_{N-1}}{4\pi^2}\langle k_{N-1}\mid
T(z_{N-1})\mid k_{N-2}>  \frac{d^2 k_{N-2}}{4\pi^2}\nonumber \\&&\times\langle k_{N-2} \mid
T(z_{N-2}
\mid k_{N-3}\rangle 
\ldots
\frac{d^2 k_3}{4\pi^2} \langle k_3\mid T(z_3)\mid k_2\rangle  \frac{d^2
k_2}{4\pi^2}\nonumber\\
&&\times \langle k_2\mid
T(z_2)\mid k_1\rangle  \frac{d^2 k_1}{4\pi^2}\langle k_1\mid T(z_1)\mid Q\rangle  .
\label{ai}
\end{eqnarray}
Here $k_1,\ldots k_{N-1}$ are the relative transverse momenta of the quark and
antiquark forming the onium. To shorten the expression, we skip the
integral symbols (and we shall continue to so doing).

At this point it is important to realize that the amplitudes entering
(\ref{ai}) depend on the longitudinal position, acquiring the eikonal
phase. So we can write
\begin{equation}
\langle k'\mid T(z) \mid k\rangle    =\e^{iz k_{\rm L}'}\langle k'\mid T \mid k\rangle  \e^{-iz k_{\rm L}},
\label{aii}
\end{equation}
where $k_{\rm L}$ and $k_{\rm L}'$ are the longitudinal momenta of the onium before and
after scattering.
\newpage
The first step is to transform the transverse momentum amplitudes\break
 $\langle k'\mid T \mid k\rangle  $ into
transverse
position amplitudes, so that we can use the explicit expression for the
onium-onium forward elastic amplitude, $T(r)$,\break derived in \cite{mu2,mu3}. We
shall use here the path integral representation, given\break by:
\begin{equation}
T(r,r_0,Y)=\pi\alpha^2rr_0\int\limits_{c-i\infty}^{c+i\infty}\frac{d\lambda}{2\pi
i}
\left(\frac{r}{r_0}\right)^{\lambda-1}\e^{{\mit\Delta}(\lambda)Y}h(\lambda),
\label{aiii}
\end{equation}
where $r$ and $r_0$ are the transverse sizes of the colliding onia. In terms of
$T(r,r_0,Y)\equiv T(r)$ we thus obtain
\begin{equation}
\langle k'\mid T(z) \mid k\rangle    =\e^{iz(k^2-k'^2)\xi}d^2\rho
\e^{i(k'-k)\rho}T(\rho)
\label{aiv}
\end{equation}
with
\begin{equation}
\xi= \frac{1}{2\nu\eta(1-\eta)}.  \label{av}
\end{equation}
In arriving at (\ref{av})
 we have used the high energy approximation for $k_{\rm L}$, {\it viz.}
\begin{equation}
k_{\rm L}= \nu - \frac{M^2}{2\nu} =\nu - \frac{k^2}{2\nu\eta(1-\eta)}.  \label{avi}
\end{equation}
 Similarly we obtain
\begin{equation}
\langle k\mid T(z) \mid Q\rangle    =\e^{iz(x_{\rm Bj}m-\xi k^2)}d^2\rho \e^{ik\rho}\langle\rho\mid T\mid
Q\rangle  ,
\label{avii}
\end{equation}
where $\langle\rho\mid T\mid Q\rangle  $ includes the photon wave function:
\begin{equation}
\langle\rho\mid T\mid Q\rangle   =T(\rho){\mit\Psi}_Q(\rho).
\label{aviia}
\end{equation}
Putting this in (\ref{ai}) and rescaling the variables
\begin{equation}
k=Q\kappa, \;\;\; \rho= \frac{r}{Q}, \;\;\; \zeta=\frac{m z}{\eta(1-\eta)}
\label{aviii}
\end{equation}
we obtain
\begin{eqnarray}
&&\hspace{-0.2cm}\langle\psi_Q\mid F_N^{\rm qel}\mid \psi_Q\rangle  =\textstyle{\frac1{
Q^2}}\e^{i(z_1-z_N)mx_{\rm Bj})}n_{\rm eff}^N\langle Q\mid T
\mid r_N\rangle  
 d^2r_N \nonumber \\
&&\hspace{-0.2cm}\times\, {\mit\Phi}[r_{N-1}-r_N,x_{\rm Bj}(\zeta_N-\zeta_{N-1})] T(r_{N-1})
 d^2r_{N-1} \nonumber \\
&&\hspace{-0.2cm}\times \,{\mit\Phi}[r_{N-2}-r_{N-1},x_{\rm Bj}(\zeta_{N-1}-\zeta_{N-2})] T(r_{N-2})
\ldots
 {\mit\Phi}[r_2-r_3,x_{\rm Bj}(\zeta_3-\zeta_2)]\nonumber \\
&&\hspace{-0.2cm}\times\, T(r_2) d^2r_2
 {\mit\Phi}[r_1-r_2,x_{\rm Bj}(\zeta_2-\zeta_1)] T(r_1) d^2r_1
\langle r_1\mid T\mid Q\rangle  ,
\label{aix}
\end{eqnarray}
with
\begin{equation}
{\mit\Phi}({\mit\Delta},a) \equiv
 \int\limits \frac{d^2 \kappa}{(2\pi)^2} \e^{i\kappa {\mit\Delta}}\e^{ia\kappa^2}=
\frac{i}{4\pi a}\e^{-i\frac{{\mit\Delta}^2}{4a}}   ,
       \label{ax}
\end{equation}
where we have assumed that $a={\rm Re}(a)+i\varepsilon$ to give a definite meaning to
the integral.

Further calculations are continued in the limit $x_{\rm Bj} \ll 1$. In this limit the
following formula can be derived by the saddle point method for $a,a'\ll 1$
\begin{eqnarray}
 &&\int\limits d^2r {\mit\Phi}(s-r,a) F(r) {\mit\Phi}(r-s',a')\nonumber \\
  &&={\mit\Phi}(s-s',a+a') F(\bar{s})
 \exp\left[i\bar{a}\left(\frac{\nabla F(\bar{s})}{F(\bar{s})}\right)^2\right],
\label{axi}
\end{eqnarray}
where
\begin{equation}
\bar{s}=\frac{a's+as'}{a+a'} ,  \label{axii}
\end{equation}
and
\begin{equation}
\bar{a} = \frac {aa'}{a+a'}   .       \label{axiia}
\end{equation}
By repeated application of this formula we finally arrive at the following
result
\begin{eqnarray}
&&\langle\psi_Q\mid F_N^{\rm qel}\mid \psi_Q\rangle   =\int\limits_0^1 d\eta\int\limits  d^2\rho
\mid{\mit\Psi}_Q(\rho,\eta)\mid^2 [n_{\rm eff}T(\rho,r_0,x_{\rm Bj})]^N \nonumber \\
&&\times \e^{-\frac{i}{2\nu}
\sum_{j=1}^N z_j\left(M_j^2-M_{j+1}^2\right)}\,,    \label{axiii}
\end{eqnarray}
where $T$ is given by (\ref{aiii})  and
\begin{equation}
M_1^2\!=\!M^2_{N+1}\!=\!-Q^2,\;\;\;
M_j^2\!=\!\left(\!\frac{N-j+1}{(\eta(1-\eta))^{1/2}}\frac{\vec{\rho}}{\rho^2}
+\vec{\nabla}\log[{\mit\Psi}(\rho)]\!\right)^2, \; j\!=\!2,\ldots,N.   \label{axv}
\end{equation}
Some details of these calculations are given in the Appendix B.

The shadow for the quasi-elastic process is now
\begin{equation}
C^{\rm qel}(A,x_{\rm Bj},Q^2)= \frac{-2}{A\sigma_t(n)}{\rm Re}\left(\sum_{N=2}^A(-1)^N
\frac{A!}{(A-N)!}\langle\langle\psi_Q\mid F_N^{\rm qel}\mid \psi_Q\rangle  \rangle  \right)\label{axvi}
\end{equation}
with the following averaging over the nuclear densities
\begin{eqnarray}
&&\langle\langle\psi_Q\mid F_N^{\rm qel}\mid \psi_Q\rangle  \rangle   =\int\limits_0^1\int\limits d\eta d^2\rho
\mid{\mit\Psi}_Q(\rho,\eta)\mid^2 [n_{\rm eff}T(\rho,r_0,x_{\rm Bj})]^N \nonumber \\
&&\times \int\limits d^2b
\int\limits_{z_N\geq z_{N-1}\geq ...\geq z_1 }  dz_N dz_{N-1}....dz_1
\rho(b,z_N)\e^{-\frac{i}{2\nu}z_N(M_N^2-M_{N+1}^2)}  \nonumber \\
&&\times\,\rho(b,z_{N-1})\e^{-\frac{i}{2\nu}z_{N-1}(M_{N-1}^2-M_{N}^2)}
\ldots\nonumber\\
&&\times\ldots
\rho(b,z_2)\e^{-\frac{i}{2\nu}z_2(M_2^2-M_{3}^2)}
\rho(b,z_{1})\e^{-\frac{i}{2\nu}z_{1}(M_{1}^2-M_{2}^2)}. \label{axvii}
\end{eqnarray}

\section{Conclusions and outlook}

We have shown that the QCD dipole picture can provide a well defined and compact
formulae for nuclear shadowing of the virtual photons. The formulae are well
suited for numerical evaluation.  They contain the following parameters:

{\it (i)} The pomeron intercept ${\mit\Delta}_P$;\vspace{0.3cm}

{\it (ii)} The nucleon size parameter $r_0$;\vspace{0.3cm}

{\it (iii)} The effective number of dipoles in the proton $n_{\rm
eff}$;\vspace{0.3cm}

{\it (iv)} The effective number of flavours (reflecting on $e_{\rm
f}^2$);\vspace{0.3cm}

{\it (v)} The scales in the elastic and in the diffractive $\gamma^*$-proton
scattering.\vspace{0.3cm}

All these parameters can, in principle, be determined from the fit of the
dipole picture to the proton data. Such a fit has been completed for the total
$\gamma^*$-proton cross-section \cite{na1}. That gave ${{\mit\Delta}}_P=.285$,
$Q_0\equiv\frac{2}{r_0}=.622$ GeV; $n_{\rm eff}e_{\rm f}^2= 3.8$;
$Y=\log\frac{1.65}{x_{\rm Bj}}$. So, as long as the fits to the diffractive
production are not available, we are left with two not fully determined
parameters: the effective number of flavours and the scale in the diffractive
dissociation.

It is important to remember that the nuclear shadowing effects are
determined by the {\it forward} diffractive amplitudes. Our formulae can
thus be employed to obtain the cross section for the forward diffractive
dissociation of the virtual photon on one nucleon. Indeed, such cross
sections on {\it one nucleon} can be obtained from the formulae of Sections 2
and 3 for $F_{N=2}^{\rm dir}$ and $F_{N=2}^{\rm qel}$:
\begin{equation}
\frac{d\sigma^{\rm
dir}}{dyd^2p_t}\big|_{p_t=0}=\frac{1}{(2\pi)^2}\langle{\mit\Psi}_Q\big|
F_{N=2}^{\rm dir} (z_1=z_2)\big|{\mit\Psi}_Q\rangle\,,
\label{A}
\end{equation}
and
\begin{equation}
\frac{d\sigma^{\rm
qel}}{d^2p_t}\big|_{p_t=0}=\frac{1}{(2\pi)^2}\langle{\mit\Psi}_Q\big|
F_{N=2}^{\rm qel} (z_1=z_2)\big|{\mit\Psi}_Q\rangle\,,
\label{B}
\end{equation}
where $p_t$ is the transverse momentum of the final proton, and $y$
defined in Eq.~(\ref{iv}) contains $M$, the diffractively produced mass.
Therefore (\ref{A}) is, in fact, a differential cross section for production
from the incident virtual photon of an object of mass $M$. On the other
hand (\ref{B}) is the total diffractive dissociation cross section {\it integrated
over diffractively excited masses}. One can resolve (\ref{B}) into the
differential cross sections $\frac{d\sigma^{\rm
qel}}{dM^2d^2_{p_t}}\!\mid_{p_t=0}$ applying a straightforward
procedure. This will be discussed elsewhere \cite{BNP}.

Our formulae, given in Sections 2 and 3 refer to two mechanisms described in the
Introduction: (a) the quasi-elastic interaction of the $q\bar{q}$ dipole and
(b) the multiple-pomeron coupling. One must remember, however, that these two
mechanisms are not mutually exclusive: they mix in the multiple scattering
terms, \ie \  there are collisions in which they both take place. Such mixed
amplitudes are also readily calculable by the methods developped in the present
paper.

Any further discussion strongly depends on the outcome of the numerical
estimates. Before this is available let us simply list the problems which,
in our opinion, seem interesting.\vspace{0.3cm}

(a) The relative importance of the multi-pomeron and the quasi-elastic
interactions. Apart from its primary interest, it is essential for an
estimation of the importance of the mixing terms.\vspace{0.3cm}

(b) An extrapolation for very small $x_{\rm Bj}$ where the unitarity is expected to
break down. It is likely that for nuclear targets this effect will occur much
earlier (\ie\  for larger $x_{\rm Bj}$) than estimated in \cite{mu3,sa1}.\vspace{0.3cm}

(c) The dependence of the kind $x_{\rm Bj}^{-{{\mit\Delta}}_P}$ which appear in BFKL
forward
amplitudes imply the existence of a large ratio of the real to imaginary part
of the forward amplitudes. We have shown already \cite{bi1,bi5} that this
influences significantly the predicted amount of shadowing. It will thus be
interesting to discuss this problem again with the present, more precise,
formulation.
\begin{center}
\section*{Appendix A}
\end{center}

In this Appendix we indicate some details of the derivation of the formula for
$n_N$, Eq.~(\ref{v}).

We start with  $n_2$.

 Using the generating
function given in \cite{mu3} and assuming that evolution of the two
exchanged pomerons goes until the rapidity $y$ is reached, we obtain the following
equation for $n_2$ ({\it c.f.} also  Eq.~(52) of \cite{mu2}):
\newpage
\begin{eqnarray}
&&\frac {dn_2(x_{01},Y,y,x,x')}{dY} \nonumber \\ &&= \frac{\alpha N_c}{\pi^2}
\int\limits_R\frac{x_{01}^2 d^2x_2}{x_{02}^2x_{12}^2} n_1(x_{12},y,x)
 n_1(x_{02},y,x') \delta(Y-y)  \nonumber \\
&&+\frac{2\alpha N_c}{\pi}\int\limits dx_{12} K(x_{01},x_{12})
n_2(x_{12},Y,y,x,x')\,,
\label{Ai}
\end{eqnarray}
where $n_1$ is the single dipole density and
\begin{equation}
K(x_{01},x_{12})= \frac1{2\pi} \int\limits_R\frac{x_{01}^2 d^2x_2}{x_{02}^2x_{12}^2}
-\delta(x_{01} -x_{12}) \log(x_{01}/\rho)          \label{Aii}
\end{equation}
is the Lipatov kernel ({\it c.f.} \cite{mu1})
which satisfies
the eigenvalue equation:
\begin{equation}
\int\limits K(x_{01},x_{12}) x_{12}^{\lambda} dx_{12} =   x_{01}^{\lambda}
\chi(\lambda)    \label{Aiii}
\end{equation}
with
$\chi(\lambda)$  given by (\ref{iii}).
We now introduce Mellin transforms to take advantage of (\ref{Aiii}):
\begin{eqnarray}
n_2(x_{01},Y,y,x,x') &=&\int\limits_{c-i\infty}^{c+i\infty} \frac{d\gamma}{2\pi i}
\tilde{n}_2(\gamma,Y,y,x,x') x_{01}^{\gamma}  \label{Aiv}\\
\tilde{n}_2(\gamma,Y,y,x,x')&=&\int\limits_0^{\infty} dx_{01} x_{01}^{-1-\gamma}
n_2(x_{01},Y,y,x,x')\,.
\label{Av}
\end{eqnarray}
Substituting (\ref{Aiv}) into (\ref{Ai}) and using (\ref{Aiii}) we obtain a
differential equation for $\tilde{n}_2$ which can be easily solved giving
\begin{equation}
\tilde{n}_2(\gamma,Y,y,x,x')=
\e^{{\mit\Delta}(\gamma)(Y-y)}g_2(\gamma,y,x,x')\,,
\label{Avi}
\end{equation}
where ${\mit\Delta}(\gamma)$ is given by (\ref{iii}) and
\begin{equation}
g_2(\gamma,y,x,x')=\frac{\alpha N_c}{\pi^2}
\int\limits_0^{\infty}dx_{01}x_{01}^{-1-\gamma}
\int\limits_R\frac{x_{01}^2 d^2x_2}{x_{02}^2x_{12}^2} n_1(x_{12},y,x)
 n_1(x_{02},y,x'). \label{Avii}
\end{equation}
Substituting (\ref{Avii}) and (\ref{Avi}) into (\ref{Aiv}) we finally have
\begin{eqnarray}
&&n_2(r,Y,y,x,x') =\frac{\alpha N_c}{\pi^2}
\int\limits_{c-i\infty}^{c+i\infty} \frac{d\gamma}{2\pi i}
r^{\gamma}\e^{{\mit\Delta}(\gamma)(Y-y)} \nonumber \\
&&\times\int\limits_0^{\infty}dx_{01}x_{01}^{-1-\gamma}
\int\limits_R\frac{x_{01}^2 d^2x_2}{x_{02}^2x_{12}^2} n_1(x_{12},y,x)
 n_1(x_{02},y,x'). \label{Aviia}
\end{eqnarray}

We now turn to $n_3$.

 Using the generating
function given in \cite{mu3} and assuming that evolution of the three
exchanged pomerons goes until the rapidity $y$ is reached, we obtain the following
equation for $n_3$:
\begin{eqnarray}
&&\frac {dn_3(x_{01},Y,y,x,x',x'')}{dY}  \nonumber \\&&= \frac{\alpha N_c}{\pi^2}
\int\limits_R\frac{x_{01}^2 d^2x_2}{x_{02}^2x_{12}^2} [n_1(x_{12},y,x)
 n_2(x_{02},Y,y,x',x'')]_{\rm sym} \delta(Y-y)  \nonumber \\
&&+\frac{2\alpha N_c}{\pi}\int\limits dx_{12} K(x_{01},x_{12}) n_3(x_{12},Y,y,x,x',x''),
\label{Axii}
\end{eqnarray}
where $n_2(x_{02},Y,y,x',x'')$  is given by (\ref{Aviia}). The symbol $_{\rm sym}$
denotes a sum of three terms, symmetrized with respect to $x,x',x''$.

Introducing Mellin transforms, as in (\ref{Aiv}), (\ref{Av}), substituting them
into
  (\ref{Ai}) and using (\ref{Aiii}) we obtain a
differential equation for $\tilde{n}_3$. The solution is found again in the
form
\begin{equation}
\tilde{n}_3(\gamma,Y,y,x,x',x'')=
\e^{{\mit\Delta}(\gamma)(Y-y)}g_3(\gamma,y,x,x',x''),
\label{Axiii}
\end{equation}
where
\begin{eqnarray}
&&g_3(\gamma,y,x,x',x'')=\frac{\alpha N_c}{\pi^2}
\int\limits_0^{\infty}dx_{01}x_{01}^{-1-\gamma}   \nonumber \\
&&\times\int\limits_R\frac{x_{01}^2 d^2x_2}{x_{02}^2x_{12}^2} [n_1(x_{12},y,x)
 n_2(x_{02},y,y,x',x'')]_{\rm sym}, \label{Axiv}
\end{eqnarray}
and $n_2(x_{02},y,y,x',x'')$ is given by (\ref{Aviia}).

The formulae for $n_4, n_5, \ldots$ are derived in a similar way.

 The integrals
over $x_1,\ldots x_N; x_1',\ldots ,x_N'$ in (\ref{v}) are performed using the formula
for the forward onium-onium amplitude $T(r,r',y)$
\begin{eqnarray}
T(r,r',y) &=& \int\limits \frac{dx}{x} \frac{dx'}{x'}\tau(x,x') n_1(r,y*,x)
n_1(r',y-y*,x')
\nonumber \\&=& \pi \alpha^2  r r' \int\limits_{c-i\infty} ^{c+i\infty}
\frac{d\gamma}{2\pi i} \e^{{\mit\Delta}(\gamma)y}\left(\frac{r}{r'}\right)^{1-\gamma}
h(\gamma)   \label{Aviib}
\end{eqnarray}
which was first derived in \cite{mu2} and \cite{na1,na2} in a sligthly more
simplified form.

Using (\ref{Aviib}) and (\ref{Aviia}), (\ref{v}) gives
\begin{eqnarray}
&&F_2^{\rm dir}(r,r_1,r_2;Y,y)=\e^{i\, (z_1-z_2\,) mx_p}
\frac{\alpha N_c}{\pi^2} r_1 r_2
\int\limits_{c-i\infty}^{c+i\infty} \frac{d\gamma}{2\pi i}
r^{\gamma}\e^{{\mit\Delta}(\gamma)(Y-y)}  \nonumber \\
&&\times\int\limits_{c-i\infty}^{c+i\infty} \frac{d\lambda_1}{2\pi i}
r_1^{\lambda_1-1} \e^{{\mit\Delta}(\lambda_1)y} h(\lambda_1)  \nonumber \\
&&\times\int\limits_{c-i\infty}^{c+i\infty} \frac{d\lambda_2}{2\pi i}
r_2^{\lambda_2-1} \e^{{\mit\Delta}(\lambda_2)y} h(\lambda_2)
{\mit\Omega}(\gamma,\lambda_1,\lambda_2)\,,       \label{Aviic}
\end{eqnarray}
where
\begin{eqnarray}
{\mit\Omega}(\gamma,\lambda_1,\lambda_2) =
\int\limits_0^{\infty}dx_{01}x_{01}^{-1-\gamma}
\int\limits_R\frac{x_{01}^2 d^2x_2}{x_{02}^2x_{12}^2}
x_{02}^{2-\lambda_1}x_{12}^{2-\lambda_2}\,,   \label{Aviid}
\end{eqnarray}
${\mit\Omega}(\gamma,\lambda_1,\lambda_2)$ can be calculated using \cite{mu1}
\begin{equation}
\frac{x_{01}^2 d^2x_2}{x_{02}^2x_{12}^2}= 2\pi x_{01}^2 \frac{dx_{02}}{x_{02}}
\frac{dx_{12}}{x_{12}} \int\limits_0^{\infty} kdk J_0(kx_{01}) J_0(kx_{02})
J_0(kx_{12}),  \label{Aix}
\end{equation}
from which the following useful identity can be obtained:
\begin{eqnarray}
\int\limits_0^{\infty}dx_{01}x_{01}^{-1-\gamma}
 \int\limits_R\frac{x_{01}^2 d^2x_2}{x_{02}^2x_{12}^2} x_{02}^{\lambda}
x_{12}^{\lambda'}= \pi
\frac{G(\lambda)G(\lambda')}{G(\gamma)}
2\pi i \delta(\gamma-\lambda-\lambda').  \label{Axa}
\end{eqnarray}
Using (\ref{Axa}) in (\ref{Aviic}) and setting $r_1=r_2=r_0$ we obtain the formula (\ref{ii}) for
$F^{\rm dir}_2$.
With the same technique one can obtain (\ref{ii}) for more than two
collisions. We indicate below the derivation for three collisions.

Substituting (\ref{Aviia}) into (\ref{Axiv}) and using (\ref{Axiii}) we obtain
\begin{eqnarray}
&&n_3(x_{01},Y,y,x,x',x'') =
\left(\frac{\alpha N_c}{\pi^2}\right)^2
\int\limits_{c-i\infty}^{c+i\infty} \frac{d\gamma}{2\pi i}
 x_{01}^{\gamma} \e^{{\mit\Delta}(\gamma)(Y-y)}
\int\limits_0^{\infty}dx_{01}x_{01}^{-1-\gamma}   \nonumber \\
&&\times\int\limits_R\frac{x_{01}^2 d^2x_2}{x_{02}^2x_{12}^2} [n_1(x_{12},y,x)
\int\limits_{c-i\infty}^{c+i\infty} \frac{d\gamma'}{2\pi i}
x_{02}^{\gamma'} \nonumber \\
&&\times\int\limits_0^{\infty}dx_{01}'(x_{01}')^{-1-\gamma'}
\int\limits_R\frac{(x_{01}')^2 d^2x_2'}{(x_{02}')^2(x_{12}')^2} n_1(x_{12}',y,x')
 n_1(x_{02}',y,x'')]_{\rm sym}. \label{Axv}
\end{eqnarray}
When (\ref{Axv}) is substituted into (\ref{v}) with $r_1=r_2=r_3=r_0$ we obtain
\begin{eqnarray}
&&F_3^{\rm dir}(r,r_0;Y,y)= \e^{i(z_1-z_N) m x_P}\nonumber \\
&&\times3\left(\frac{\alpha N_c}{\pi^2}\right)^2
\int\limits_{c-i\infty}^{c+i\infty} \frac{d\gamma}{2\pi i}
 x_{01}^{\gamma} \e^{{\mit\Delta}(\gamma)(Y-y)}
\int\limits_0^{\infty}dx_{01}x_{01}^{-1-\gamma}   \nonumber \\
&&\times\int\limits_R\frac{x_{01}^2 d^2x_2}{x_{02}^2x_{12}^2} T(x_{12},r_0,y)
\int\limits_{c-i\infty}^{c+i\infty} \frac{d\gamma'}{2\pi i}
x_{02}^{\gamma'} \nonumber \\
&&\times\int\limits_0^{\infty}dx_{01}'(x_{01}')^{-1-\gamma'}
\int\limits_R\frac{(x_{01}')^2 d^2x_2'}{(x_{02}')^2(x_{12}')^2} T(x_{12}',r_0,y)
 T(x_{02}',r_0,y)\,, \label{Axvv}
\end{eqnarray}
where $T(r,r',y)$ is the dipole--dipole forward scattering amplitude given\break by~(\ref{Aviib}).

The integrals over $dx_{01}$,$d^2x_2$ and $dx_{01}'$ $d^2x_2'$ can be
 performed using
again the identity (\ref{Axa}) and taking into account the fact that the
dependence of $T(r,r',y)$ on $r$ and $r'$ is in the form of a power law.
The result of these operations is the Eq.~(\ref{ii}) for $N=3$.

\begin{center}
\section*{Appendix B}
\end{center}

In this Appendix we give a few intermediate steps of the calculations of
Section 3 which lead to Eq.~(\ref{axiii}).

First, by repeated application of (\ref{axi}) we obtain
\begin{eqnarray}
&&\hspace{-0.2cm}\langle\psi_Q\mid F_N^{\rm qel}\mid \psi_Q\rangle  =n_{\rm eff}^N Q^{-2}\e^{i(z_1-z_N)mx_{\rm Bj})}\langle Q\mid T \mid
r_N\rangle  
 d^2r_N \nonumber \\ &&\hspace{-0.2cm}\times {\mit\Phi}(r_{1}-r_N,x_{\rm Bj}(\zeta_N-\zeta_{N-1}))
S(r_N,r_1;\zeta_N,\ldots ,\zeta_1)d^2 r_1\langle r_1\mid T\mid Q\rangle,
\label{Bi}
\end{eqnarray}
where
\begin{eqnarray}
&&S(r_N,r_1;\zeta_N,\ldots ,\zeta_1) = \prod_{j=2}^{N-1}\frac{d\lambda_j}{2\pi i}
 {\mit\Lambda}(\lambda_j,r_0,Y) r_j^{\lambda_j}\nonumber \\&&\times \exp \left(
i\frac{a_j\lambda_j}
{r_j^2} +i 2\sum_{j,k=2}^{N-1} \lambda_j \lambda_k \frac{\vec{r}_j \vec{r}_k}
{r_j^2r_k^2} a_{jk}\right)
\label{Bii}
\end{eqnarray}
with
\begin{eqnarray}
\vec{r}_j&=& \frac{(z_N-z_j)\vec{r}_1 +(z_j-z_1)\vec{r}_N}{z_n-z_1},
\label{Biii}\\
a_j&=& \frac{(z_N-z_j)(z_j-z_1)}{z_n-z_1}, \label{Biv}\\
a_{jk}&=& \frac{(z_N-z_j)(z_k-z_1)}{z_n-z_1}. \label{Biva}\\
{\mit\Lambda}(\lambda,r_0,Y)&=& n_{\rm eff}\pi\alpha^2r_0^{2-\lambda}
\e^{{\mit\Delta}(\lambda)Y}h(\lambda)     \label{Bv}
\end{eqnarray}
is taken from the integrand of Eq.~(\ref{aiii}) and $n_{\rm eff}$ added for the
reasons explained in the main text.

We finally perform the integral over $d^2r_N$, again in the saddle point
approximation which is valid for small $x_{\rm Bj}$. Then
${\mit\Phi}(r_1-r_N,x_{\rm Bj}(\zeta_N-\zeta_1))$ is very strongly peaked around $r_n
\approx r_1$, so that in leading order all $r_j$'s become $r_1$. This
approximation puts a new kind of factor, $\nabla\psi_Q/\psi_Q$, into the over
all phase of the expression and we obtain (returning to $z_j$'s and $\rho_j$'s)
\begin{eqnarray}
&&\langle\psi_Q\mid F_N^{\rm qel}\mid \psi_Q\rangle  = n_{\rm eff}^N\e^{ix_{\rm Bj}(z_1-z_N)}
\int\limits_0^1d\eta \int\limits d^2\rho
\mid\psi_Q(\rho,\eta)\mid^2 \nonumber \\&&\times \exp
\Bigg[\frac{i}{2\nu}\Bigg[\frac1{\eta(1-\eta)\rho^2}\sum_{j=2}^N[2(N-j)+1](z_j-z_1)
\nonumber \\&&+ 2\sum_{j=2}^N (z_j-z_1)\frac{\vec{\rho}\cdot\vec{\nabla}\psi_Q(\rho)}
{[(\eta(1-\eta)]^{1/2} \rho^2 \psi_Q(\rho)} +(z_n-z_1)\frac{[\vec\nabla
\psi_Q(\rho)]^2}{\psi_Q(\rho)^2}\Bigg]\Bigg]\nonumber \\
&&\times \prod_{j=1}^{N}\int\limits \frac{d\lambda_j}{2\pi i}
 {\mit\Lambda}(\lambda_j,r_0,Y) \rho^{\lambda_j}\,.
\label{Bvi}
\end{eqnarray}
One can check by a direct substitution of $M_j^2$ given by (\ref{axv}) into
(\ref{axiii}) that it reduces to (\ref{Bvi}). The version of
 $\langle\psi_Q\mid F_N^{\rm qel}\mid \psi_Q\rangle $ given in (\ref{axv}) has, however, a
clearer physical interpretation.

\bigskip

We thank A.~Capella, A.~Kaidalov, A.~Krzywicki and R.~Peschanski for helpful
discussions. Part of this work was done when the authors visited LPTHE Orsay
and CE Saclay. We thank D.~Schiff and J.~Zinn--Justin for the kind hospitality.
This work was supported in part by the KBN Grant No 2 PO3B 083 08 and by PECO
grant from the EEC Programme ``Human Capital Mobility'', Network ``Physics at High
Energy Colliders'' , Contract No ERBICIPDCT 940613.

\end{document}